\documentclass[
 amsmath,amssymb,
 aps,superscriptaddress,notitlepage,nofootinbib,prl, twocolumn]{revtex4-1}

\usepackage{tcolorbox}
\usepackage{graphicx}
\usepackage[utf8]{inputenc}
\usepackage{latexsym}
\usepackage{multirow,booktabs}
\usepackage{amsfonts,amsmath,amssymb,amssymb,amsthm}
\usepackage{hyperref}
\hypersetup{colorlinks=false,pdfborder={0 0 0}}
\usepackage[english]{babel}

\usepackage{graphicx,color,array}
\usepackage{times}

\usepackage{verbatim}

\usepackage{changes}

\newcommand{\ba}{\begin{eqnarray}}
\newcommand{\be}{\begin{equation}}
\newcommand{\ee}{\end{equation}}

\newcommand{\bn}{\begin{equation*}}
\newcommand{\en}{\end{equation*}}

\newcommand{\ea}{\end{eqnarray}}
\newcommand{\ban}{\begin{eqnarray*}}
\newcommand{\ean}{\end{eqnarray*}}
\newcommand{\Tr}{\operatorname{tr}}
\newcommand{\tr}{\operatorname{tr}}

\newcommand{\sket}[1]{{\ensuremath{\lvert#1\rangle}}}
\newcommand{\lket}[1]{{\ensuremath{\left\lvert#1\right\rangle}}}
\newcommand{\ket}[1]{\if@display\lket{#1}\else\sket{#1}\fi}

\newcommand{\sbra}[1]{{\ensuremath{\langle#1\rvert}}}
\newcommand{\lbra}[1]{{\ensuremath{\left\langle#1\right\rvert}}}
\newcommand{\bra}[1]{\if@display\lbra{#1}\else\sbra{#1}\fi}

\newcommand{\sbraket}[2]{{\ensuremath{\langle#1\rvert#2\rangle}}}
\newcommand{\lbraket}[2]{{\ensuremath{\left\langle#1\!\left\rvert\vphantom{#1}#2\right.\!\right\rangle}}}
\newcommand{\braket}[2]{\if@display\lbraket{#1}{#2}\else\sbraket{#1}{#2}\fi}

\newcommand{\sketbra}[2]{{\ensuremath{\lvert #1\rangle\!\langle #2\rvert}}}
\newcommand{\lketbra}[2]{{\ensuremath{\left\lvert #1\right\rangle\!\!\left\langle #2\right\rvert}}}
\newcommand{\ketbra}[2]{\if@display\lketbra{#1}{#2}\else\sketbra{#1}{#2}\fi}

\newcommand{\proj}[1]{\ketbra{#1}{#1}}

\newcommand{\tp}{\otimes}
\newcommand{\idd}{\openone}

\newcommand{\tx}[1]{\textnormal{#1}}

\newcommand{\sotimes}{\scalebox{0.7}{$\otimes\,$}}

\newcommand{\rA}{\mathrm{A}}

\newcommand{\rB}{\mathrm{B}}

\newcommand{\rC}{\mathrm{C}}
\newcommand{\rD}{\mathrm{D}}

\newcommand{\M}{\mathsf{M}}
\newcommand{\X}{\mathsf{X}}
\newcommand{\Y}{\mathsf{Y}}
\newcommand{\Z}{\mathsf{Z}}
\newcommand{\D}{\mathsf{D}}

\newtheorem{theorem}{Theorem}
\newtheorem{lemma}[theorem]{Lemma}

\begin{document}

\title{Device-independent entanglement certification of all entangled states}

\author{Joseph Bowles}
\affiliation{ICFO-Institut de Ciencies Fotoniques, The Barcelona Institute of Science and Technology, 08860 Castelldefels (Barcelona), Spain}
\author{Ivan \v{S}upi{\'c}}
\affiliation{ICFO-Institut de Ciencies Fotoniques, The Barcelona Institute of Science and Technology, 08860 Castelldefels (Barcelona), Spain}
\author{Daniel Cavalcanti}
\affiliation{ICFO-Institut de Ciencies Fotoniques, The Barcelona Institute of Science and Technology, 08860 Castelldefels (Barcelona), Spain}
\author{Antonio Ac{\'i}n}
\affiliation{ICFO-Institut de Ciencies Fotoniques, The Barcelona Institute of Science and Technology, 08860 Castelldefels (Barcelona), Spain}
\affiliation{ICREA-Instituci\'o Catalana de Recerca i Estudis Avan\c cats, Pg. Lluis Companys 23, Barcelona, 08010, Spain}

\date{\today}

\begin{abstract}
We present a method to certify the entanglement of all bipartite entangled quantum states in a device-independent way. This is achieved by placing the state in a quantum network and constructing a correlation inequality based on an entanglement witness for the state. Our method is device-independent, in the sense that entanglement can be certified from the observed statistics alone, under minimal assumptions on the underlying physics. Conceptually, our results borrow ideas from the field of self-testing to bring the recently introduced measurement-device-independent entanglement witnesses into the fully device-independent regime.
\end{abstract}

\maketitle
\emph{Introduction}--- The certification of entanglement is a vital task in quantum information processing for which much effort has been put into developing optimal methods \cite{GuhneTothReview}. Typically, one uses an approach based on entanglement witnesses \cite{entanglement}; since every entangled state violates a suitably chosen entanglement witness, one can in principle certify the entanglement of any entangled state. This approach however requires the precise knowledge of the measurements performed during the certification. At best, this means that much effort has to be put into the characterisation of the experimental setup and sources of error must be known and accounted for. At worst, if the system under investigation is highly complex or poorly understood, the method may not be applicable or a false positive certification may result \cite{Rosset_FalsePositive}.

\begin{figure}
\includegraphics[scale=1]{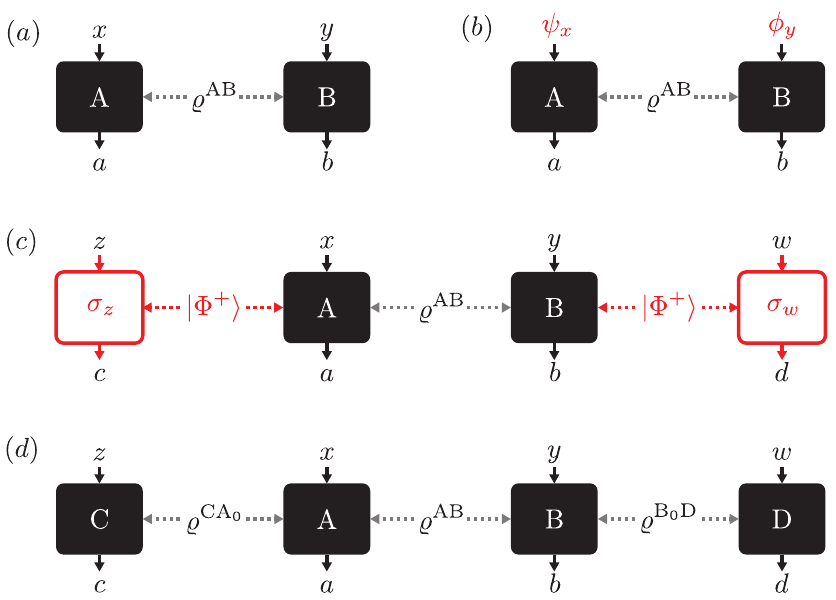}
\caption{\label{scenario}Scenarios for entanglement certification. Red denotes trusted states/devices. {\bf (a)} Standard Bell scenario for device independent entanglement certification. The estimated probabilities $p(ab|xy)$ are tested for violation of a Bell inequality in order to certify the entanglement of the state $\varrho^{\rA\rB}$. {\bf (b)} Scenario for MDI entanglement certification. Here, the inputs are given by trusted quantum states $\psi_{x}$ and $\phi_{y}$. {\bf(c)} Equivalent MDI scenario in which the inputting of the states $\psi_{x}$ and $\phi_{y}$ in scenario (b) is replaced by giving Alice and Bob each one half of a maximally entangled state and performing local measurements on them. {\bf(d)} Our proposal for DI entanglement certification.  The entangled state $\varrho^{\rA\rB}$ to be detected is placed in a network containing additional auxiliary entangled states. Using self-testing techniques, these entangled states are certified to be maximally entangled and perform the expected measurements as required in (c).}
\end{figure}

A solution to this problem recently came from the field of device-independent (DI) quantum information \cite{bellreview,DIQKD,DIdimension,DIrandomness}. Here, the aim is to certify physical properties of quantum systems without requiring precise knowledge of the underlying physics, that is, by treating all devices as black boxes processing classical information. In the case of entanglement certification, one requires that the state under investigation violates a Bell inequality \cite{bell,bellreview}, a linear function of the observed experimental probabilities which is bounded for all separable states. Since the Bell inequality is a function of the observed probabilities only and independent of the specific physical realisation, entanglement can be certified without any assumptions on the performed measurements, making this approach practically attractive. 

The advantages of this approach however come at a price: not all entangled states are capable of violating a Bell inequality \cite{werner89,Barrett_2002,TothAcin06,Almeida07,Wiseman07,Augusiak_2014,bowles16,algoconstruction,algoconstruction2}. For example, the two-qubit isotropic state
\begin{align}
\varrho(p)=p\ketbra{\Phi^{\text{+}}}{\Phi^{\text{+}}}+(1-p)\frac{\openone}{4}
\end{align}
where $\ket{\Phi^{\text{+}}}=(\ket{00}+\ket{11})/\sqrt{2}$, is entangled for $p\geq1/3$, however Bell inequality violation with projective measurements is impossible if $p\lesssim 0.68$ \cite{Grot2017} ($p \lesssim 0.45$ in the case of general measurements \cite{Grot2017,POVMsimulation}). For a large class of states, a device-independent entanglement detection method based on the violation of a standard bipartite Bell inequality therefore cannot be used. 

This naturally leads to the question of whether the entanglement of all entangled states can be certified device-independently using an alternative approach. In this work we show the answer to be yes by considering networks of quantum states. Network scenarios have already been shown to be useful for DI entanglement certification, through the phenomenon of activation of Bell nonlocality \cite{activation1,activation2,activation3}. In the present work we propose a method of entanglement certification where the state under investigation is placed in a network featuring two additional bipartite auxiliary states. 
The certification of entanglement is achieved via the violation of a correlation inequality based on an entanglement witness for the state and borrows ideas from the fields of self-testing, semi-quantum games and measurement-device-independent (MDI) entanglement witnesses \cite{buscemi,MDIEW1,CavalcantiHallWiseman,MDIEW2}. Moreover, our construction is fully DI, requiring knowledge of the observed statistics only. %

\emph{Previous work}--- In the standard scenario for DI entanglement certification, two parties, Alice and Bob, share a bipartite quantum state $\varrho^{AB}$ and wish to ensure that it is entangled. As mentioned, one way to achieve this is via a Bell test, in which each party treats his/her subsystem as a black box on which they perform a number of possible measurements labeled by the classical variables $x$ for Alice and $y$ for Bob, obtaining outcomes $a$ and $b$ respectively (see Fig.\ \ref{scenario} (a)). At the end of the experiment, they estimate the joint probabilities $p(a,b\vert x,y)$ of obtaining outcomes $a$ and $b$ for measurements $x$ and $y$. A DI certification of entanglement is a proof of the entanglement of $\varrho^{AB}$ which follows from the probabilities $p(a,b\vert x,y)$ alone, i.e.\ without requiring assumptions about the specific physical system under investigation or the form of the measurement operators. This is equivalent to proving that the probabilities $p(a,b\vert x,y)$ cannot be produced by any separable state, and can be achieved via Bell inequality violation; since separable states always produce local statistics, Bell inequality violation certifies the entanglement of the state $\varrho^{AB}$. 

As noted, there exist entangled states which do not violate any Bell inequality \cite{werner89,Barrett_2002,TothAcin06,Almeida07,Wiseman07,Augusiak_2014,bowles16,algoconstruction,algoconstruction2}. Hence, the entanglement of many states cannot be certified in this scenario. One partial solution to this problem came in the form of MDI entanglement witnesses (MDIEWs) \cite{MDIEW1,MDIEW2}. Here, the Bell test scenario is modified so that the measurement inputs are given by quantum states $\psi_x$ and $\phi_y$, as opposed to the classical labels $x$ and $y$ (see Fig.\ \ref{scenario} (b)). In the general construction, the set of quantum inputs for each party should be informationally complete on the local Hilbert spaces of the state under investigation. With this, a Bell-like correlation inequality can be constructed from every entanglement witness and the entanglement of all entangled states can be certified. 

However, this approach is not DI since it assumes the knowledge of the input states. In what follows we show how one can remove this assumption and achieve a fully DI certification for all entangled states. Here we concentrate on the case of two-qubit systems for the sake of simplicity. Generalisations to higher dimensions and multipartite states will be discussed in a later section and more in detail in a longer, technical version of this work \cite{PRA}. For two-qubit states, a convenient choice of a tomographically complete set of states to use in an MDIEW protocol are the eigenstates of the Pauli matrices, i.e. 
\begin{align}\label{pauliproj} 
\{\psi_{x}\}=\{\phi_{y}\}=\{\ket{0},\ket{1},\ket{+},\ket{-},\ket{R},\ket{L}\},
\end{align}
being $\ket{+/-}=\frac{1}{\sqrt{2}}(\ket{0}\pm\ket{1})$ and $\ket{R/L}=\frac{1}{\sqrt{2}}(\ket{0}\pm i\ket{1})$. 
Our starting point is to see that the inputting of the states $\psi_x$ and $\phi_y$ is mathematically equivalent to the following (see Fig.\ \ref{scenario}(c)). Prepare two ancilla states both in the state $\ket{\Phi^{\text{+}}}$ and give one qubit of each to Alice and to Bob. On the remaining two qubits, perform one of the three Pauli measurements, specified by $z=1,2,3$ and $w=1,2,3$. Conditioned on the choice of Pauli measurements and the corresponding outcomes, Alice and Bob's qubits are projected in one of the states in \eqref{pauliproj}. 
This replacement is not DI, as it still assumes the form of the maximally entangled states and measurements on them. However, it is possible to use self-testing techniques to achieve a DI certification of these states and measurements~\cite{McKMosca,POVMrand}. The main idea of our protocol is to incorporate these self-testing techniques into the MDI protocol for entanglement detection and promote it into a fully DI protocol that detects any entangled state.

\emph{DI entanglement certification in networks}--- We are now ready to define our scenario. We extend the standard Bell scenario to involve two more parties, Charlie and Daisy (see Fig.\ \ref{scenario}(c)). As before, the aim is to certify the entanglement of the state $\varrho^{AB}$ shared between Alice and Bob, however we now introduce two auxiliary states, $\varrho^{\rC\rA_0}$ shared between Charlie and Alice and $\varrho^{B_0D}$ shared between Bob and Daisy. Denoting the set of linear operators on Hilbert space $\mathcal{H}$ by $\mathcal{B}(\mathcal{H})$ we have $\varrho^{AB}\in\mathcal{B}(\mathcal{H}_{\rA}\otimes\mathcal{H}_{\rB})$, $\varrho^{\rC\rA_0}\in\mathcal{B}(\mathcal{H}_{\rC}\otimes\mathcal{H}_{\rA_0})$ and $\varrho^{B_0D}\in\mathcal{B}(\mathcal{H}_{\rB_0}\otimes\mathcal{H}_{\rD})$. We work in a DI scenario in the sense that we assume (i) the validity of quantum theory but not the precise form of the states and measurements and (ii) that the network of Fig.\ \ref{scenario} (c) correctly describes the experimental setup. Note that since we are only interested in certifying the entanglement of $\varrho^{AB}$, no restrictions are placed on the states $\varrho^{CA_0}$ and $\varrho^{B_0D}$, in particular they may (and indeed will) be entangled. We now move to the central result of our work.
\begin{tcolorbox}[title=Main result,title filled]
The entanglement of any entangled state $\varrho^{\rA\rB}$ can be certified in the scenario of Fig.\ \ref{scenario}(c) as follows: \\[2pt]
(i) The parties perform local measurements on their subsystems to obtain the statistics $p(c, a, b, d|z, x, y, w)$. \\ [4pt]
(ii) The following is then verified: \\[2pt]
\emph{Self-testing}-- The marginal distributions $p(c,a\vert z,x)$ and $p(b,d\vert y,w)$ maximally violate a Bell inequality that certifies that the auxiliary states each contain a maximally entangled state and that Charlie and Daisy each perform Pauli measurements on their subsystems. \\[2pt]
\emph{Entanglement certification}-- The correlations violate an additional inequality $\mathcal{I}[p(c, a, b, d|z, x, y, w)]\geq 0$ that certifies $\varrho^{\rA\rB}$ is entangled.
\end{tcolorbox}
 
Let us first discuss step (i), considering two-qubit systems. Charlie and Daisy have a choice of three measurements $z,w=1,2,3$ with outcomes $c,d=\pm 1$. Alice and Bob have a choice of seven measurements $x,y=1,2,3,4,5,6,\star$ with outcomes $a,b=\pm1$. The auxiliary states are chosen to be $\varrho^{\rC\rA_0}=\varrho^{\rB_0\rD}=\proj{\Phi^{\tx{+}}}$. Charlie's measurements are given by the three Pauli observables $\sigma_{\tx{z}}, \sigma_{\tx{x}}, \sigma_{\tx{y}}$,  for $z=1,2,3$. Alice's measurements for the inputs $x=1,\cdots, 6$ are given by the rotated Pauli observables $(\sigma_{\tx{z}}\pm\sigma_{\tx{x}})/\sqrt{2}$, $(\sigma_{\tx{z}}\pm\sigma_{\tx{y}})/\sqrt{2}$, $(\sigma_{\tx{x}}\pm\sigma_{\tx{y}})/\sqrt{2}$ acting on the $\mathcal{H}_{\rA_0}$ space. For the input $x=\star$, Alice's measurement is given by $\{\proj{\Phi^{\tx{+}}},\openone-\proj{\Phi^{\tx{+}}}\}$ acting on the joint space $\mathcal{H}_{\rA_0}\otimes\mathcal{H}_{\rA}$. Measurements for Bob and Daisy are chosen analogously. 

The Bell inequality we use for our self-testing in step (ii) of the protocol is as follows (here we focus on Charlie and Alice). Denote by $E_{x,y}$ the expectation value of the measurements $x$ and $y$. Consider the Bell inequality 
\begin{multline}\label{chainedCHSH}
\mathcal{J}= E_{1,1}+E_{1,2}+E_{2,1}-E_{2,2}  \\
+E_{1,3}+E_{1,4}-E_{3,3}+E_{3,4}  \\
+E_{2,5}+E_{2,6}-E_{3,5}+E_{3,6} \leq 6.
\end{multline}
This bound follows from the fact that each line of the above is a CHSH Bell inequality \cite{CHSH}, each upper bounded by 2. Using the state $\varrho^{\rC\rA_0}$ and measurements described above one achieves a maximal violation of each CHSH inequality and so $\mathcal{J}=6\sqrt{2}$. Note that each of Charlie's measurements appears in exactly two of the lines. Since the maximum violation of a single CHSH inequality requires two anti-commuting measurements \cite{popescu92,Braunstein92,robustsinglet}, one would expect that the maximum violation of \eqref{chainedCHSH} require three anti-commuting measurements for Charlie. This is indeed the case, as described in the following lemma (see \cite{note} for related results).

\begin{lemma}\label{lemma1}
Let Charlie and Alice share the state $\ket{\psi}\in\mathcal{H}_\rC\otimes\mathcal{H}_{\rA_0}$ and  denote by $\Z^{\rC}$, $\X^{\rC}$, $\Y^{\rC}$ three $\;\pm1\;$ outcome observables for Charlie. If one observes a Bell inequality violation of $\mathcal{J}=6\sqrt{2}$ then there exist local auxiliary states $\ket{00}\in [\mathcal{H}_{\rC''}\otimes\mathcal{H}_{\rC'}] \otimes [\mathcal{H}_{\rA_0''}\otimes\mathcal{H}_{\rA_0'}]$ and a local unitary $U=U_\rC\otimes U_\rA$ such that
\begin{align} \label{ststate}
U[\ket{\psi}\otimes\ket{00}] &= \ket{\xi}\tp \ket{\Phi^{\tx{+}}}^{\rC'\rA'},\\ \label{stx}
U[\X^\rC\ket{\psi}\otimes\ket{00}] &= \ket{\xi} \tp \sigma_{\tx{x}}^{\rC'}\ket{\Phi^{\tx{+}}}^{\rC'\rA'},\\ \label{stz}
U[\Z^\rC\ket{\psi}\otimes\ket{00}] &= \ket{\xi}\tp \sigma_{\tx{z}}^{\rC'}\ket{\Phi^{\tx{+}}}^{\rC'\rA'},\\ \label{sty}
U[\Y^\rC\ket{\psi}\otimes\ket{00}] &= \sigma_{\tx{z}}^{\rC''}\ket{\xi}\tp \sigma_{\tx{y}}^{\rC'}\ket{\Phi^{\tx{+}}}^{\rC'\rA'}, 
\end{align}
where $\ket{\xi}$ takes the form 
\begin{align}\label{junk}
\ket{\xi} =\ket{\xi_0}^{\rC\rA}\otimes\ket{00}^{\rC''\rA''} + \ket{\xi_1}^{\rC\rA}\otimes\ket{11}^{\rC''\rA''}. 
\end{align}
\end{lemma}
Here we use superscript to denote the Hilbert space on which an operator acts nontrivially. For example $\X^\rC\ket{\psi} \equiv (\X^\rC\tp \openone^{\rA_0})\ket{\psi}$. The above lemma can be understood as follows. The observation $\mathcal{J}=6\sqrt{2}$ implies that the state $\ket{\psi}$ must contain a two-qubit maximally entangled subspace and that two of Charlie's measurements must be given by the observables $\sigma_{\tx{x}}$ and  $\sigma_{\tx{z}}$ in this space (equations \eqref{ststate} to \eqref{stz}). From \eqref{sty}, the third measurement of Charlie is equivalent to first measuring the observable $\sigma_{\tx{z}}$ on the state $\ket{\xi}$, and then measuring either $\sigma_{\tx{y}}$ or $-\sigma_{\tx{y}}$ on his half of the maximally entangled state depending on this first outcome. We can therefore understand the above as Charlie measuring either $\{\sigma_{\tx{x}},\sigma_{\tx{y}},\sigma_{\tx{z}}\}$ or $\{\sigma_{\tx{x}},-\sigma_{\tx{y}},\sigma_{\tx{z}}\}$ on the maximally entangled state, with some unknown probability that depends on the precise (unknown) form of $\ket{\xi}$. This reflects the fact that the only two non-unitarily equivalent sets of mutually anti-commuting measurements on a qubit are given by $\{\sigma_{\tx{x}},\pm\sigma_{\tx{y}},\sigma_{\tx{z}}\}$, which are related via transposition (or equivalently complex conjugation) in the computational basis. A full proof of Lemma 1 can be found in \cite{PRA}.

Strictly speaking we have not self-tested the three Pauli measurements on the maximally entangled state due to the additional $\sigma_\tx{z}$ measurement in \eqref{sty}. However, this does not prevent us from using the MDIEW technique. The intuitive reason for this is as follows. Since the measurements $\{\sigma_{\tx{x}},\sigma_{\tx{y}},\sigma_{\tx{z}}\}$ and $\{\sigma_{\tx{x}},-\sigma_{\tx{y}},\sigma_{\tx{z}}\}$ are related via transposition, the states that Alice receives for the input to the MDIEW protocol (see \eqref{pauliproj}) are essentially either $\psi_x$ or $\psi_x^T$ with some unknown probability. Using transposed quantum inputs $\psi_x^T$ for Alice in a MDIEW protocol with a product state $\varrho^{\rA\rB}=\sigma_{\rA}\otimes\sigma_{\rB}$ is mathematically equivalent to using the standard inputs $\psi_x$ on the state $\sigma_{\rA}^T\otimes\sigma_{\rB}$. However, since this state remains a separable, this cannot lead to false positive entanglement detection. 

We now move to entanglement certification part of step (ii) of the protocol. Fix an entangled two-qubit quantum state $\tilde{\varrho}^{\rA\rB}$ for which to perform the entanglement certification. The correlation inequalities we consider are constructed from an entanglement witness for the state $\tilde{\varrho}^{\rA\rB}$ and are inspired from those found in \cite{buscemi,MDIEW1,MDIEW2}. For every entangled $\tilde{\varrho}^{\rA\rB}$ there exists a Hermitian linear operator $\mathcal{W}$, called an entanglement witness, such that $\tr(\mathcal{W}\varrho^{\rA\rB}) \geq 0$ for every separable state $\varrho^{\rA\rB}$ and $\tr(\mathcal{W}\tilde{\varrho}^{\rA\rB}) < 0$. Consider the projectors $\pi_{c|j}=\frac{1}{2}[\openone+c\,\sigma_{j}]$ with $c=\pm1$ and $j=1,2,3$, that is, projectors onto the plus and minus eigenspaces of the Pauli observables. Since these form a basis of the set of Hermitian matrices, any entanglement witness for a two-qubit state may be decomposed as 
\begin{equation}
\label{witness}
\mathcal{W} = \sum_{cdzw}\omega_{cd}^{zw}\,\pi_{c\vert z}\otimes \pi_{d\vert w}.
\end{equation}
The inequality we consider is then
\begin{equation}
\label{ineq}
\mathcal{I}=\sum_{cdzw}\omega_{cd}^{zw}\;p(c,+,+,d\vert z,x=\star,y=\star,w)\geq 0,
\end{equation}
which is satisfied if $\varrho^{\rA\rB}$ is a separable state, however can be violated using $\tilde{\varrho}^{\rA\rB}$. To see this, write the probabilities arising from the network of Fig.\ \ref{scenario}(c) as
\small\begin{align}
&p(c,+,+,d\vert z,x=\star,y=\star,w) \\
&\quad=\Tr\left[\M^{\rC}_{c|z}\otimes \M^{\rA_0\rA}_{+|\star} \otimes \M^{\rB\rB_0}_{+|\star} \otimes \M^{\D}_{d|w} \;\varrho^{\rC\rA_0}\otimes\varrho^{\rA\rB}\otimes \varrho^{\rB_0\rD}\right],\nonumber
\end{align}\normalsize
where the $\M_{i\vert j}$ are the local measurement operators. Since there are no restrictions on the auxiliary states or measurements, we may assume that these states are pure and the measurements $\M^{\rC}_{c|u}$ and $\M^{\D}_{d|w}$ projective without loss of generality. We may therefore write
\small\begin{align}
&p(c,+,+,d\vert z,x=\star,y=\star,w) \\
&\;=\Tr\left[\openone\otimes \M^{\rA_0\rA}_{+|\star} \otimes \M^{\rB\rB_0}_{+|\star} \otimes \openone \;\proj{\psi}_{c|z}^{\rC\rA_0}\otimes\varrho^{\rA\rB}\otimes \proj{\psi}_{d|w}^{\rB_0\rD}\right],\nonumber
\end{align}\normalsize
where $\ket{\psi}_{c|z}=\M^{\rC}_{c|z}\ket{\psi}^{\rC\rA_0}$ and $\ket{\psi}_{d|w}=\M^{\rD}_{d|w}\ket{\psi}^{\rB_0\rD}$. From step (ii), we may use Lemma \ref{lemma1} to replace the auxiliary states and measurements in the above, e.g.\ $\ket{\psi}_{c|z}=U^\dagger[\ket{\xi}\otimes\pi^{\rC'}_{c|z}\ket{\Phi^{\tx{+}}}]$ for $z=1,2$. After some work (see \cite{PRA} Supp.\ Mat.\ F for details) one obtains
\begin{align}
\mathcal{I}=\tr\left[\mathcal{W}\;\Lambda(\varrho^{\rA\rB})\right]
\end{align}
where $\Lambda(\cdot)$ can be shown to be a local positive map on all separable states. One thus has that $\Lambda(\varrho^{\rA\rB})$ is separable if $\varrho^{\rA\rB}$ is separable and so $\mathcal{I}\geq 0$ for all separable $\varrho^{\rA\rB}$. The proof of this follows the same structure as the MDIEW technique, however one must take a bit more care due to the additional complications implied by Lemma 1. 

It remains to show that one can violate $\mathcal{I}$ using the state $\tilde{\varrho}^{\rA\rB}$. First generate auxiliary states $\varrho^{\rC\rA_0}=\varrho^{\rB_0\rD}=\proj{\Phi^{\tx{+}}}$ and perform the measurements detailed in step (i) so that the self-testing conditions of step (ii) are satisfied. One then has
\begin{align}
&p(c,+,+,d\vert z,x=\star,y=\star,w)= \\ 
&\Tr\left[\pi_{c|z}\sotimes\proj{\Phi^{\tx{+}}}\sotimes\proj{\Phi^{\tx{+}}}\sotimes\pi_{d|w}\, \proj{\Phi^{\tx{+}}}\sotimes \tilde{\varrho}^{\rA\rB}\sotimes\proj{\Phi^{\tx{+}}}\right] \nonumber\\
&\quad\quad=\frac{1}{4}\tr\left[\proj{\Phi^{\tx{+}}}\sotimes\proj{\Phi^{\tx{+}}}\, \pi_{c|z}^T\sotimes \tilde{\varrho}^{\rA\rB} \sotimes \pi_{d|w}^T\right] \\
&\quad\quad=\frac{1}{16}\tr\left[\pi_{c|z}\otimes \pi_{d|w} \,\tilde{\varrho}^{\rA\rB} \right], 
\end{align}
where we have used $\tr_{\rA}[\proj{\Phi^{\tx{+}}}\, \pi^{\rA}_{i|j}\sotimes \idd]=\frac{1}{2}\pi_{i|j}^T$ in the second and third line. One thus has 
\begin{align}\label{viol}
\mathcal{I}&=\frac{1}{16}\sum_{czdw}\omega_{cd}^{zw}\tr[\pi_{c|z}\otimes \pi_{d|w} \,\tilde{\varrho}^{\rA\rB}] \\
\mathcal{I}&=\frac{1}{16}\Tr[\mathcal{W}\tilde{\varrho}^{\rA\rB} ]<0,
\end{align}
hence certifying the entanglement of $\tilde{\varrho}^{\rA\rB}$. 

\emph{High dimension and multipartite states}--- Our method can be used to certify the entanglement of bipartite states of any dimension. Every bipartite entangled state of dimension $d\times d$ violates an entanglement witness of the form 
\begin{align}
\mathcal{W}=\sum_{ij}\omega_{ij}\pi_i\otimes\pi_j,
\end{align}
where the set $\{\pi_i\}$ consists of (at least) $d^2$ linearly independent quantum states. As in the qubit case states $\{\pi_i\}$ can be prepared in a device independent manner by distant parties Charlie and Daisy which now share with Alice and Bob respectively a tensor product of $N$ maximally entangled pairs of qubits, where $N = \lceil\log{d}\rceil$. Specifically, by performing a parallel self-test of Lemma \ref{lemma1}, one can certify tensor products of the Pauli measurements for Charlie and Daisy which provide an informationally complete set of states $\{\pi_i\}$ for Alice and Bob; see \cite{PRA}. 

The same idea can also be utilised to certify the presence of entanglement in multipartite states of any dimension. Each party would share a suitable maximally entangled state with an auxiliary party, which is used to self-test the preparation of an informationally complete set of states. We stress however that this approach is not suitable to detect genuine multipartite entanglement. This is because the set of $k-$separable states is not closed under partial transposition on individual parties, so the imprecision in the sign of the self-tested $\pm \sigma_y$ measurement may lead to false positive results. 

\emph{Noise robustness}--- 
It is important to ask whether our protocol can be made robust to noise. Suppose that the violations of the Bell inequality \eqref{chainedCHSH} in step (ii) of the protocol differ from the maximum value. Since self-testing protocols are robust, the observed violation guarantees that the states and measurements must be close, though not exactly equal to the desired ones. In particular, suppose equations \eqref{stx} -- \eqref{sty} hold up to a small value $\theta$ in the $\ell_2$ norm, i.e.
\begin{align}
\vert\vert U[\X^\rC\ket{\psi}\otimes\ket{00}] - \ket{\xi} \tp \sigma_{\tx{x}}^{\rC'}\ket{\Phi^{\tx{+}}}^{\rC'\rA'} \vert\vert \leq \theta,
\end{align}
and similarly for equations \eqref{stz}, \eqref{sty}. In \cite{PRA} we show that entanglement can still be certified if one changes the bound of \eqref{ineq} to read $\mathcal{I}\geq -\mathcal{O}(\theta)$. As a result, for non-maximal violations, a fraction of entangled states close to the separable states is no longer detected.  

\emph{Discussion}--- A number of improvements to the self-testing part of out protocol would strengthen our results. For example, it may be possible to lower the requirement on the number of inputs/outputs by self-testing more efficient sets of informationally complete measurements in high dimension (e.g. by using mutually unbiased bases or symmetric positive operator valued measures). Additionally, the overall noise robustness of the entanglement certification would benefit from improvements to the robustness of self-testing statements, which at the moment are typically weak. In principle, our technique can also be applied to convex sets of bipartite quantum states other than the separable set, provided that the set be closed under local unitaries and local transpositions. Furthermore, one may be able to apply our general method to other DI tasks such as quantum key distribution and randomness certification where MDI protocols already exist \cite{MDIQKD,MDIrand}. 

To conclude, our work opens new perspectives for entanglement certification by connecting different concepts such as self-testing and MDI protocols in a quantum network. For weakly entangled states where optimal Bell inequalities are not known, our method provides a general construction that is easily applicable to all states. Furthermore, it allows for DI entanglement certification of entangled states admitting a so-called local hidden variable model for which the standard approach fails. We hope that the present results motivate further studies on DI protocols that could be boosted by the use of quantum networks.

\section{acknowledgements}
We are thankful for useful discussions with Paul Skrzypczyk, Nicolas Brunner, Marco T{\'u}lio Quintino, Flavien Hirsch, Thomas Vidick, Matteo Lostaglio,  Micha\l{} Oszmaniec and Alexia Salavrakos. This work was supported by the Ram\'on y Cajal fellowship, Spanish MINECO (QIBEQI FIS2016-80773-P and Severo Ochoa SEV-2015-0522), the AXA Chair in Quantum Information Science, Generalitat de Catalunya (CERCA Programme), Fundaci\'{o} Privada Cellex and ERC CoG QITBOX.

\bibliography{BowlesNL}

\begin{thebibliography}{38}
\bibitem [{not()}]{note}%
  \BibitemOpen
  \href@noop {} {\ }\bibinfo {note} {Note that McKague and Mosca
  \cite{McKMosca} showed an equivalent result to Lemma 1 based on the
  Mayers-Yao self-test (see also \cite{POVMrand} and \cite{CabelloElegant}). In
  \cite{PRA}, we extend these results (see also \cite{leash}) by providing a
  generalization of Lemma 1 to a parallel self-test in arbitrary dimension as
  well as an analysis of the effect of imperfect statistics.}\BibitemShut
  {Stop}%
\end{thebibliography}

\end{document}